\documentclass[%
reprint,
superscriptaddress,
 amsmath,amssymb,
aps,
 prb,
]{revtex4-2}

\usepackage[unicode=true,colorlinks=true,citecolor=blue,urlcolor=blue]{hyperref}
\usepackage[english]{babel} 
\usepackage{amssymb}
\usepackage{amsmath}
\usepackage{txfonts}
\usepackage{mathdots}
\usepackage[classicReIm]{kpfonts}
\usepackage{graphicx}
\usepackage[dvipsnames]{xcolor}
\def\arrvline{\hfil\kern\arraycolsep\vline\kern-\arraycolsep\hfilneg}
\usepackage{dcolumn}
\usepackage{bm}
\usepackage{multirow}
\usepackage[normalem]{ulem}
\usepackage{tabularx}
\usepackage{comment}
\usepackage{booktabs}
\usepackage{array} 

\newcommand{\e}{\mathrm{e}}

\let\ifr\i
\renewcommand{\i}{{\rm i}}

\begin{document}
\preprint{APS/123-QED}

\title{Towards determination of  intrinsic spin properties in MAPbI$_3$ crystals using spin noise spectroscopy}

\author{M.~Vladimirova}
\affiliation{Laboratoire Charles Coulomb, UMR 5221 CNRS/Universit\'{e} de Montpellier, France}
\author{S.~Cronenberger}
\affiliation{Laboratoire Charles Coulomb, UMR 5221 CNRS/Universit\'{e} de Montpellier, France}
\author{L.~Doyennette}
\affiliation{Laboratoire Charles Coulomb, UMR 5221 CNRS/Universit\'{e} de Montpellier, France}
\author{D.~R.~Yakovlev}
\affiliation{Experimentelle Physik 2 Technische Universit\"{a}t Dortmund,   Dortmund, Germany}
\author{B.~Turedi}
\affiliation{Laboratory of Inorganic Chemistry, Department of Chemistry and Applied Biosciences, ETH Z\"{u}rich, CH-8093 Z\"{u}rich, Switzerland}
\affiliation{Laboratory for Thin Films and Photovoltaics, Empa---Swiss Federal Laboratories for Materials Science and Technology, CH-8600 D\"{u}bendorf, Switzerland}
\author{M.~V.~Kovalenko}
\affiliation{Laboratory of Inorganic Chemistry, Department of Chemistry and Applied Biosciences, ETH Z\"{u}rich, CH-8093 Z\"{u}rich, Switzerland}
\affiliation{Laboratory for Thin Films and Photovoltaics, Empa---Swiss Federal Laboratories for Materials Science and Technology, CH-8600 D\"{u}bendorf, Switzerland}

\begin{abstract}

Spin noise spectroscopy in back- and forward-scattering geometries is used to study the spin dynamics of charge carriers in a MAPbI$_3$ (MA = methylammonium) single crystal. 
The spectra reveal electrons and holes with different degrees of localization.
At low temperatures ($T \approx 4$~K), hole spin relaxation is dominated by the hyperfine interaction with nuclear spins.
 The broadening of the spectra with increasing temperature can be attributed to thermally activated hopping of localized holes.
  From the thermal activation energies, we estimate hole localization lengths that are consistent with both the hyperfine-induced dephasing rates and the zero-field frequency shifts measured at low temperature.
   By contrast, the electron spin dynamics cannot be explained by the hyperfine interaction alone: even at the lowest excitation powers, it is affected by the probe light, and we observe a slowing down of spin relaxation with increasing power.
   Comparison of the back- and forward-scattered spin noise signals indicates that the spatial distribution of carriers is inhomogeneous: electrons are located preferentially near the crystal surface, weakly localized holes (characterized by slower spin relaxation) are found mostly away from the surface, and strongly localized holes are somewhere in between.
\end{abstract}

\date{\today}
\maketitle
\section{Introduction}
Methylammonium  lead triiodide (MAPbI$_3$)  is one of the most studied 
hybrid  organic-inorganic  lead halide perovskites. 
The great attention that it has attracted was initially due to its 
photovoltaic efficiency \cite{BerryAdvEnergyMat2022}. 
Later on 
MAPbI$_3$ has been reported to exhibit up to tens of nanoseconds long 
carrier spin lifetimes, positioning it as an attractive system for 
exploring spin coherence in a solution-processable semiconductor 
\cite{OdenthalNatPhys2017, kirsteinSpinDynamicsElectrons2022,
KoturHyperfinePRB2026,garcia-arellanoEnergyTuningElectronic2021,
lagueHyperfineInducedAsymmetricOverhauser2026,KozlovSpinNoiseHalide2025,
kirsteinResonantSpinAmplification2025,
KoptevaOpticalOrientationExcitons2025}. 
However, despite extensive efforts, the nature of carriers and 
thus the mechanisms governing spin relaxation in these systems remain 
insufficiently understood. 

\begin{figure}
	\includegraphics[width=3.4in]{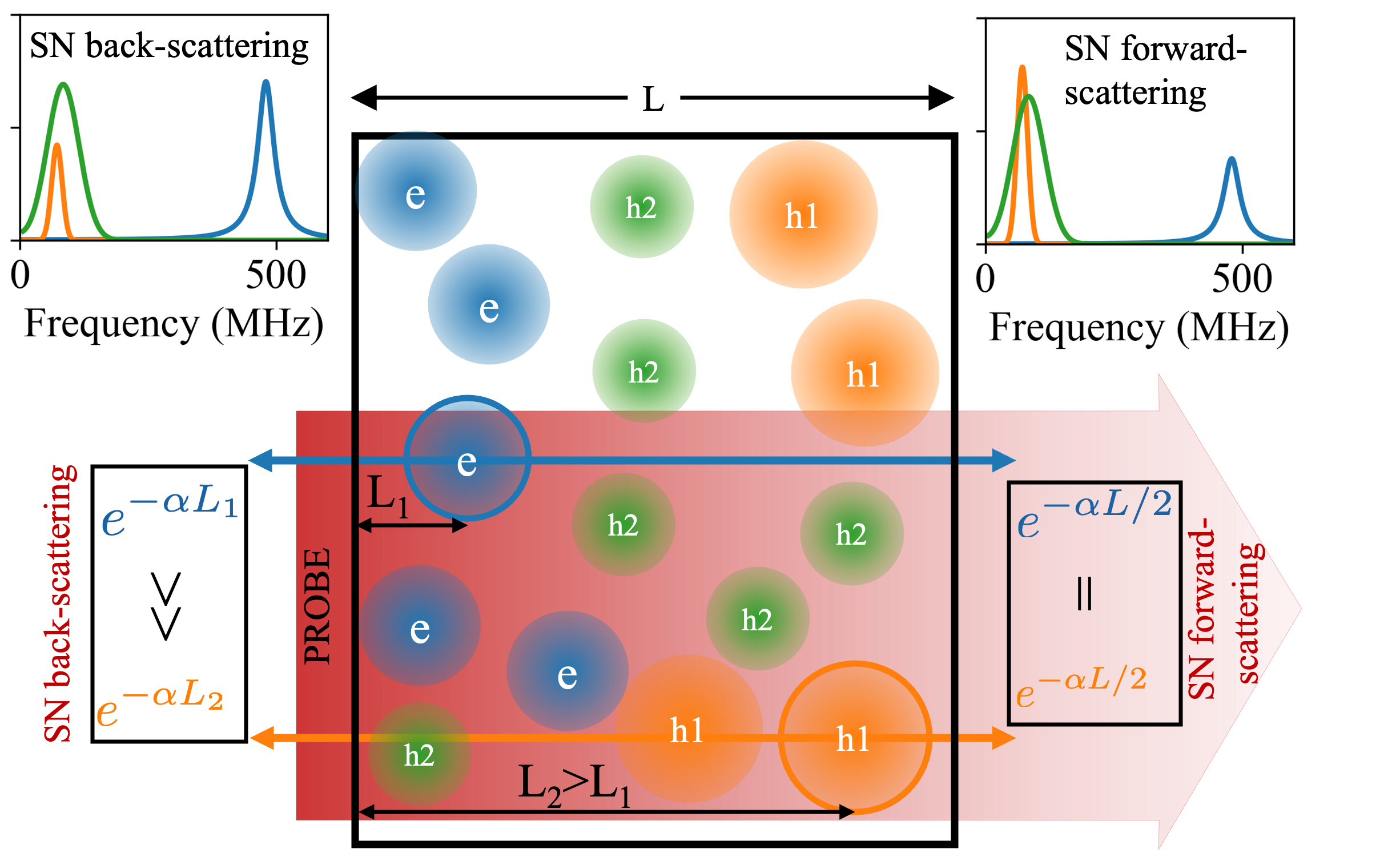}
	\caption{  
		Schematic of two experimental geometries and of the effect of probe absorption on the difference between SN spectra in forward- and back scattering geometries. 
		In forward scattering geometry (right side) the SN power is independent on the spin position with respect to the surface.  In back-scattering geometry (left side) the spins situated deep in the sample contribute less (e. g. h$_1$ at distance $L_2$), than those situated close to the surface (e.g. electron at distance $L_1$). 
		By analyzing relative spectral areas,  the  carrier distribution in the sample   
		(electrons in blue, two hole components in orange ($h_1$) 
		and green ($h_2$) ), is deduced from their relative contributions 
		in back- and forward-scattering SN spectra. The latter are shown schematically 
		on the left and right sides of the sample, respectively, 
		with the same color code.  }
	\label{fig:Fig1}
\end{figure}

A central open question for this material is the microscopic 
mechanism that limits carrier spin relaxation in the orthorhombic, 
low-temperature phase. 
In conventional semiconductors two mechanisms are usually invoked. 
For free (delocalized) carriers, spin relaxation is expected to proceed via the Dyakonov-Perel (DP) mechanism, in which momentum scattering randomizes the spin  precession  in a  $k$-dependent effective spin–orbit field associated with local inversion-symmetry breaking  \cite{DyakonovPerel1972}. While some experiments point out that DP spin-relaxation mechanism  is inactive in MAPbI$_3$ crystals with orthorhombic symmetry \cite{KoptevaOpticalOrientationExcitons2025}, other studies report Rashba-type spin splittings at low temperature in hybrid perovskites, attributed to low-symmetry domains, or local electric fields from defects \cite{NiesnerStructuralFluctuations,lafalceRashbaSplittingOrganic2022}. Whether such extrinsic or local inversion-symmetry breaking is strong enough to make DP relaxation relevant in MAPbI$_3$ crystals remains an open question.

 For localized carriers, by contrast, spin dephasing is 
 expected to be dominated by the hyperfine (HF) interaction 
 with the surrounding bath of nuclear spins, since a spatially 
 confined carrier wavefunction couples to a finite, 
 fluctuating ensemble of  nuclear spins, whose random Overhauser 
 field drives dephasing at zero external magnetic field \cite{Merkulov02}. 
 In lead halide perovskites HF coupling is much stronger for 
 holes than for electrons 
 \cite{lagueHyperfineInducedAsymmetricOverhauser2026,
 KoturHyperfinePRB2026,koturResonantCoolingNuclear2026}, 
 but  for both types of carriers the resulting coherence times 
 depend on the localization volume and eventually on the spin 
  hopping rate \cite{Gribakin2025,Glazov_hopping}. 
 To make the picture even more complex,  
 the presence of holes with different localization volumes 
 (referred to as { weakly (h$_1$) and strongly (h$_2$)}
 localized holes) has been reported \cite{KoturHyperfinePRB2026}. 
 Remarkably, weakly localized holes have only been observed  
 in single crystals of MAPbI$_3$, but  not in thin polycrystalline  films 
 \cite{garcia-arellanoEnergyTuningElectronic2021,
 lagueHyperfineInducedAsymmetricOverhauser2026}.

\begin{figure*}
	\includegraphics[width=6.8in]{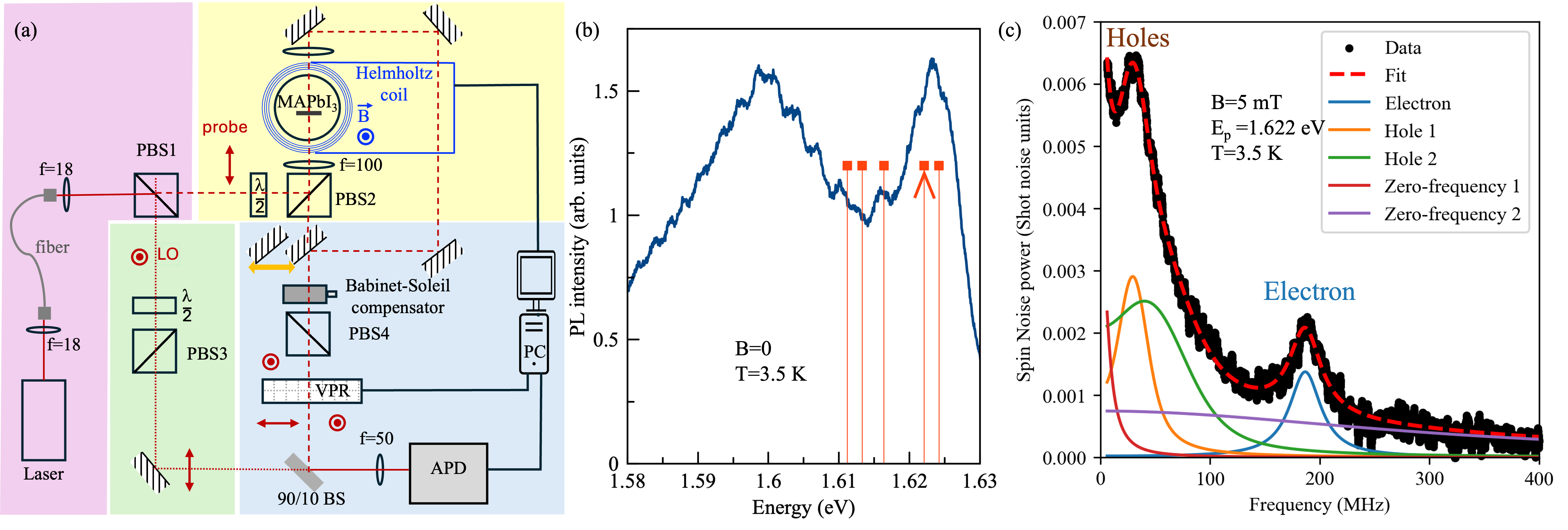}
	\caption{
	(a) Sketch of the experimental setup. Four colour-coded regions:
	the laser excitation part (magenta) up 
	to the splitting into LO (green) and probe (yellow). 
	 The detection part, which includes mixing  of the probe light, 
	 Raman-scattered by the sample with the LO, is shown in blue. 
	Legend: PBS $\equiv$ Polarizing beam splitter; 
	 BS $\equiv$ Beam sampler; VPR $\equiv$  
	 Variable phase retarder; $\lambda/2$ $\equiv$ Half wave plate; 
	 APD  $\equiv$ avalanche photodiode. 
	 Focal lengths are given in millimetres. 
	 (b) PL spectrum measured  at $B=0$, $T=3.5$~K, under excitation at $2.33$~eV.
	 Symbols and vertical lines indicate energies of the probe beam, $E_p$, in SN measurements, 
	 arrow points to the probe energy corresponding to  the spectrum in (c). 
	 (c)   SN power spectrum (black symbols) measured in 
	 forward geometry at $B=5$~mT, $T=3.5$~K, 
	 probe energy $E_p=1.622$~eV and  power density $P=350$~W/cm$^2$. 
	 Red dashed line shows the fit to the multi-component 
	 function with five distinct components shown by solid lines, 
	 including electron (blue Voigt profiles), 
	 two hole components (orange ($h_1$) and green ($h_2$) 
	 Voigt profiles) and two zero-frequency components 
	 (red and lilac Lorentzians).
	}
	\label{fig:setup}
\end{figure*}

Spin noise 
(SN) spectroscopy is  a powerful technique which has been extensively used to 
explore intrinsic spin properties of  conventional bulk semiconductors \cite{oestreich_spin_2005,Gribakin2025}.
It relies on
 mapping of stochastic 
fluctuations of electron and/or hole spins at thermal 
equilibrium onto Kerr (Faraday) rotation and/or ellipticity 
fluctuations of the Raman-scattered light using continuous 
wave probe beam tuned below exciton 
resonance~\cite{oestreich_spin_2005,Smirnov-review}.
Fourier spectrum  of these fluctuations provides new 
insights into  spin dynamics of electrons and holes, as 
compared to photo-induced Kerr (Faraday) rotation, 
and  {photoluminescence-based Hanle} and polarization recovery experiments,
in particular due to minimized density of photocarriers. 
Moreover, because the SN spectra of electrons and holes in MAPbI$_3$ at non-zero magnetic field
are centered at different frequencies, their individual contributions can be easily separated, in contrast to Hanle  and polarization recovery experiments,  where the magnetic field is scanned through zero
and additional hypotheses are needed to separate electrons and holes \cite{KoturHyperfinePRB2026,lagueHyperfineInducedAsymmetricOverhauser2026}.

SN experiments in perovskites has been reported only recently \cite{KozlovSpinNoiseHalide2025,kozlovSpinNoiseReveals2025}.
These first experiments,  performed in forward-scattering geometry, 
identified a single  Lorentzian SN resonance in  MAPbI$_3$  thick crystals.
It  was assigned to  {delocalized holes}, with a spin relaxation 
time of several nanoseconds that decreases with increasing temperature, 
 a trend the authors argued was inconsistent with a 
HF-limited mechanism and instead attributed to DP 
relaxation of delocalized carriers. 
This assignment — a different carrier population 
(delocalized rather than localized) relaxing via a different mechanism 
(DP rather than HF) — stands in direct tension with the 
picture built from reflection-geometry photoluminescence (PL) and time-resolved  Kerr rotation  measurements 
\cite{KoptevaOpticalOrientationExcitons2025,KoturHyperfinePRB2026,
kirsteinSpinDynamicsElectrons2022}. 
The discrepancy has been noted by the authors of Ref.~\onlinecite{KozlovSpinNoiseHalide2025} as possibly arising 
from the difference between forward and back-scattering geometries, 
i.e., from the possibility that each technique preferentially probes 
a different subset of the carrier population.

Our results bridge, at least partly, the gap between 
the observations made using 
these different experimental techniques.
Relying on comparison of SN spectra with the Hanle  and  polarization recovery 
profiles measured  via PL in similar samples \cite{KoturHyperfinePRB2026}, 
we  suggest that these methods probe the same families of spin carriers.  
In addition, SN spectra detected at different energies in both forward- and 
back-scattering geometries indicate that  spatial 
distribution of carriers is inhomogeneous.
Electrons are more abundant close to the sample surface,
 weakly localized holes are mainly present 
in the bulk of the sample, 
while strongly localized holes are situated between them, 
as sketched  Fig.~\ref{fig:Fig1}.

The paper is organized as follows. 
In  the next section MAPbI$_3$ sample and SN spectroscopy setup 
are described.
This is followed by the presentation of a typical SN spectrum 
and identification of its constituents 
in Section~\ref{sec:Identification}. 
Probe energy, power and temperature dependencies are analyzed in 
Sections \ref{sec:probe_energy}, \ref{sec:power} and \ref{sec:temp}, 
respectively. 
Section \ref{sec:forwardback} shows how a combination 
of back- and forward-scattering SN spectra allows us to uncover 
spatial inhomogeneity of the spin carrier distribution. 
Finally, in Section \ref{sec:concl} we summarize the results  and 
discuss our experimental 
observations  in comparison  with available 
data from the literature.
Section   \ref{sec:concl-concl}  presents
the conclusions.

\section{Sample and experimental technique}
\label{sec:sample}
We investigate a single crystal of MAPbI$_3$ with a thickness 
of 15.9 $\mu$m, prepared using the inverse temperature 
crystallization technique \cite{Chen2019,Alsalloum2020,Nazarenko2017}.
Optical and spin properties of similar samples have been reported in Refs.~\cite{KoptevaOpticalOrientationExcitons2025,KoturHyperfinePRB2026,koturResonantCoolingNuclear2026}.
It was established that at cryogenic temperatures the sample 
presents orthorhombic structure and free exciton energy 
$E_X=1.636$~eV, while the maximum of PL is observed at lower energy,
as shown in Fig.~\ref{fig:setup}(b). 
The latter originates from  recombination of spatially separated, 
localized electrons and holes. 

The sample is fixed on a glass substrate and  is placed on a cold finger 
in a variable 
temperature closed-cycle cryostat surrounded by a Helmholtz coil. 
The resulting magnetic field is oriented  in the plane of the sample,  
along $[100]$ axis (normal to the optical axis).

The SN setup used in this work is presented in 
Fig.~\ref{fig:setup}(a). 
The experiments are based on the homodyne detection of SN 
\cite{RSI2016,Petrov2018,sterinTemperaturedependentElectronSpin2022,Kamenskii2020,
Cronenberger2019,Gribakin2025,zibinskiySpinNoiseLocalized2026}.
This choice is motivated by the possibility to 
acquire SN spectra at fixed magnetic field, 
and using probe powers significantly lower 
than in traditional self-homodyne detection \cite{KozlovSpinNoiseHalide2025}. 
The excitation part of the setup (magenta) starts from a 
continuous-wave, low noise, tunable around $1.6$~eV  external cavity diode laser, TOPTICA DL pro.
The laser beam  is sent through a mono-mode, polarization-conserving  
optical fiber onto a polarized beam splitter (PBS). 
The PBS splits the laser beam into two parts: the yellow part 
corresponds to the probe beam, and the green one to the local oscillator (LO). 
The power of the beams is then independently adjustable via 
dedicated half-wave plates and polarizers.  
A beam sampler (BS) directs a small part ($<$10~\%) of 
the vertically polarized probe normally to sample surface. 
The beam is focused by the $100$~mm focal length  lens, 
yielding a Gaussian spot with a full width at half maximum (FWHM) of 
$21$~µm on the sample surface.

 We then analyze either reflected (back-scattering) or transmitted (forward-scattering) 
 light (blue part of the setup).
 While the probe beam is polarized vertically, the SN signal is encoded in the horizontally polarized spin-flip Raman-scattered light.
 In order to detect the corresponding signal, 
 the reflected (transmitted) probe beam is filtered out with $1/1000$ 
 extinction ratio by a Babinet-Soleil compensator and a polarizer.
The  remaining scattered light is modulated by variable 
phase retarder (VPR), mixed with the LO in a $90/10$ beam 
splitter and  sent to an avalanche photodiode (APD).
The resulting signal is digitized and Fourier transformed by an AC240 acquisition card.
 
A typical spectrum measured in forward-scattering geometry at 
probe energy $E_p=1.622$~eV, magnetic field $B=5$~mT, 
temperature $T=3.5$~K and power density $P=350$~W/cm$^2$ 
is shown in  Fig.~\ref{fig:setup}(c). 
In the next section we focus on the identification of various 
contributions to this SN spectrum.

\section{Identification of electron and hole contributions to spin noise  spectra}
\label{sec:Identification}
\begin{figure}
	\includegraphics[width=3.3in]{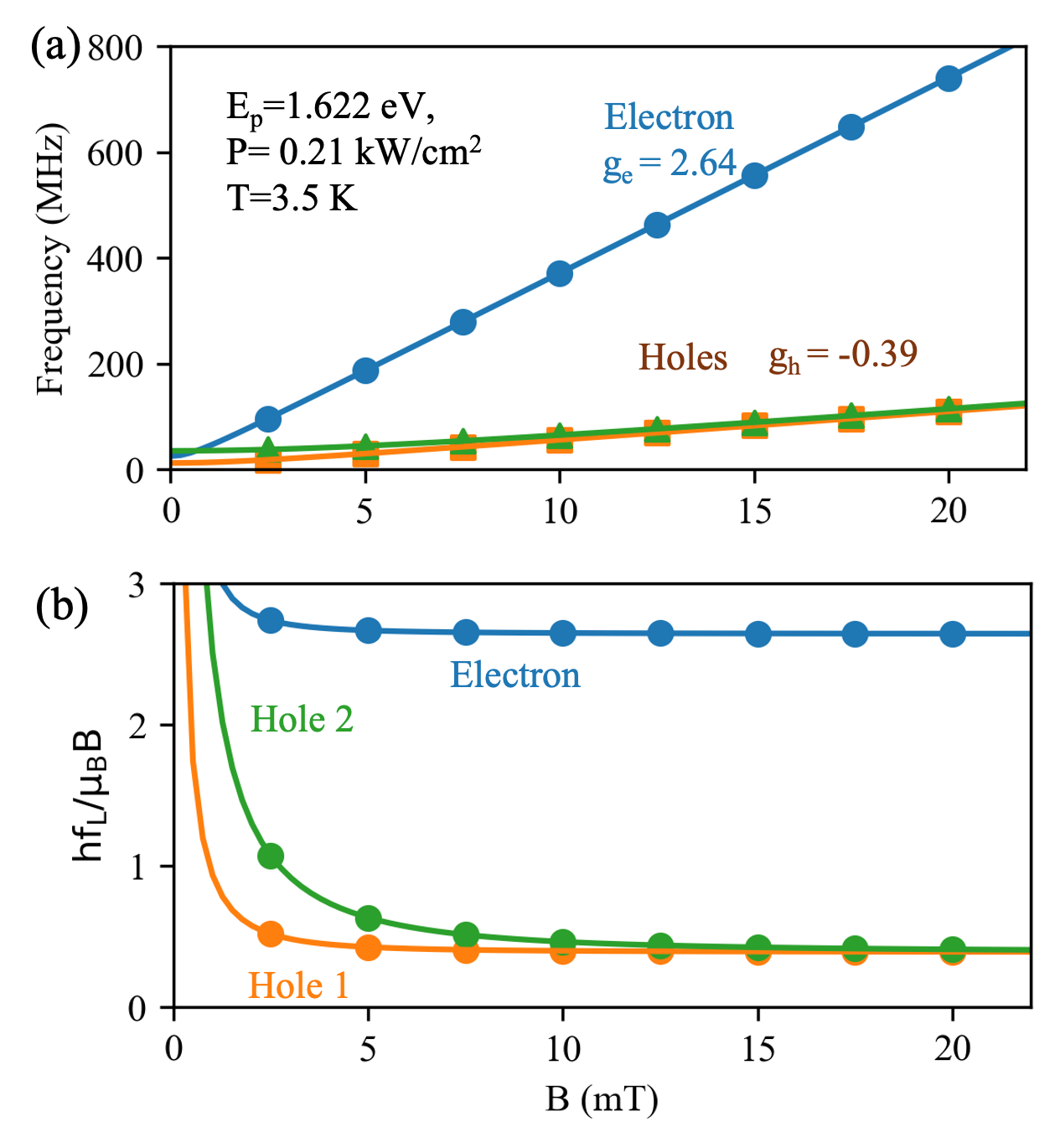}
	\caption{ Frequencies of the spectral peaks 
	extracted from the fits of the SN spectra measured 
	at various magnetic fields.  
	$E_p=1.622$~eV, $P=210$~W/cm$^2$, $T=3.5$~K.
        Solid lines are fits  after Eq.~\eqref{eq:fL}.
	}
	\label{fig:Fig_B_dep}
\end{figure}
Two main peaks at non-zero frequencies can be 
seen  in Fig.~\ref{fig:setup}(c). 
On the basis of electron and hole  $g$-factor values reported 
in the literature, $g_e= 2.46$ to $2.98$ and $g_h=-0.28$ to $-0.71$ \cite{Kirstein2022NatComm}, we attribute the peak at
$190$~MHz to electron spin fluctuations, and the peak at $30$~MHz to the hole ones.

In most physical situations, spin fluctuations can be described by Voigt function, that is 
a convolution of the Gaussian and Lorentzian shapes \cite{Smirnov-review}. 
Moreover, to accurately 
describe the experimentally measured spectra, 
 two zero-frequency  contributions with Lorentzian shape, 
 need to be taken into account 
 (red and lilac curves in Fig.~\ref{fig:setup}(c)).
 These contributions do not depend on the magnetic field,
but are power and detection energy dependent. Their origin is not very clear, 
they may be related to the exciton dephasing, longitudinal component of spin relaxation, generation-recombination noise, and some uncontrolled 
instrumental response.

The spectrum of hole spin fluctuations 
needs at least two Voigt functions with different width and frequencies to be reproduced. 
This is consistent with  findings of Kotur {\it et al.}, who analyzed the shape of 
Hanle  {and} polarization recovery curves, 
and deduced from this analysis the presence of holes with different 
localization energies {\cite{KoturHyperfinePRB2026}}.
The resulting fit and its constituents are shown in 
Fig.~\ref{fig:setup}(c) by dashed and solid lines, respectively. 
Remarkably, at $B=5$~mT two hole components (orange, h$_1$, and green, h$_2$) 
have not only different widths but also
different Larmor frequencies.

Figure~\ref{fig:Fig_B_dep} shows the
magnetic field dependence of these Larmor frequencies. 
One can see that the Larmor frequency at the lowest fields   deviates  
from linear dependence determined by $g$-factor.
To illustrate this effect better, we show  in Fig.~\ref{fig:Fig_B_dep}(b)
how the $g$-factor deviates from the constant value 
by dividing the Larmor frequency  by  $\mu_BB/h$, where $\mu_B$ is Bohr magneton, 
$h$ is Planck constant. 
This deviation is a characteristic fingerprint of the HF interaction of the localized charge carriers with the randomly oriented spins of the lattice nuclei~\cite{Glazov_Ivchenko_2012}. 
HF field adds to the applied magnetic field, shifting up the 
corresponding frequency, so that the field dependence of 
Larmor frequency 
can be modeled by:
\begin{equation}
f_L=\sqrt{f_{0}^2+(g_{e(h)}\mu_B B/h)^2} , 
\label{eq:fL}
\end{equation}
see solid lines in Figs.~\ref{fig:Fig_B_dep}(a, b). Here $g_h=-0.39$   and  $g_e=2.64$ are hole and electron $g$-factors, respectively,   and $f_{0}$ is the HF-induced 
shift equal to $25\pm10$~MHz for electrons, $12\pm5$~MHz for holes with narrow 
spectrum (h$_1$, orange line) and $35\pm10$~MHz for 
holes with broad spectrum (h$_2$, green line), see also  Table~\ref{tab:f0_FWHM}. 
Note, that we assign negative 
value to $g_h$ and positive to $g_e$ on the basis of the theoretical 
considerations \cite{Kirstein2022NatComm,lagueSpinCoherenceRelaxation2024}
and assume that the field dependence is characterized by the 
same $g$-factor for both hole contributions.

These   observations suggest that  all the carriers contributing 
to the SN signal are  localized at $T=3.5$~K, and that
their spin dephasing is due to fluctuations of the 
nuclear spins within the corresponding localization volume.   
The values of $f_0$ are determined by both the HF 
coupling strength and the localization volume.
Therefore,  the localization is stronger 
for h$_2$ holes with broad spectral line and large zero-field 
frequency shift, than for h$_1$  holes, for which this shift is only
detectable at lowest magnetic field, see Fig.~\ref{fig:Fig_B_dep}(b). 
For electrons in MAPbI$_3$  the HF interaction has been 
shown to be weaker than for holes~\cite{lagueHyperfineInducedAsymmetricOverhauser2026,KoturHyperfinePRB2026}. 
However, we do observe a  deviation of  dependence $hf_L/\mu_BB$ from 
constant, at least at $B=2.5$~mT, see Fig.~\ref{fig:Fig_B_dep}(b).

Within the model of localized carriers, the zero-field 
frequency $f_0$  of each spectral component 
is determined by the  standard deviation  of the spin precession frequency components in the random Overhauser field. 
 It can be shown that  Gaussian parameter of the SN Voigt profile, $\sigma_0$,
is  given by ~\cite{Glazov_Ivchenko_2012}
\begin{equation}
	 \sigma_0=f_0/\sqrt{2}. 
	 \label{sigma_0}
\end{equation}
This means that  FWHM 
 	of Voigt profile and $\sigma_0$ are related.
However, FWHM may  contain other contributions, namely the spin relaxation rate, which is unrelated to   HF interaction. 
It corresponds to the Lorentzian 
component of the Voigt profile, characterized by the parameter $\gamma$, 
and governs the relaxation of the spin component parallel to the sum of the nuclear and external magnetic fields~\cite{Glazov_Ivchenko_2012}. 
Specifically, FWHM  of Voigt profile can be approximated as~\cite{oliveroEmpiricalFitsVoigt1977}
\begin{equation}
	\mathrm{FWHM}=1.07\,\gamma +\sqrt{0.87\,\gamma ^2+5.5\sigma_0^2}.
	\label{eq:FWHM}
\end{equation}
Because at $B=0$ it is impossible to reliably separate different carrier contributions, 
and since no significant variation of FWHM with magnetic field is observed, 
for the analysis presented in 
the set of data shown in Fig.~\ref{fig:Fig_B_dep} we  use  $\sigma_0$ 
defined by Eq.~(\ref{sigma_0}). 
The corresponding value of  FWHM  is then calculated  from  the average value of $\gamma$
using Eq.~\eqref{eq:FWHM}.
Finally, the spin relaxation time $T_2^*$ is related to FWHM as:
\begin{equation}
	T_2^*=1/(\pi\,\mathrm{FWHM}),
	\label{eq:T2}
\end{equation}
and within our model it includes both reversible (dephasing) and irreversible spin relaxation.
 The values of $f_0$, FWHM and $T_2^*$  corresponding to 
 Fig.~\ref{fig:Fig_B_dep} are given in Table~\ref{tab:f0_FWHM}. 

\begin{table}[h]
  \centering
  \caption{ Evaluated parameters of MAPbI$_3$ crystal.  $g$-factors and $f_0$ values are  obtained from fits of Eq.~\eqref{eq:fL} to the field dependence data 
    ({\it cf} Fig.~\ref{fig:Fig_B_dep}). The corresponding $\sigma_0$ values (Eq.~\eqref{sigma_0}) are kept fixed in  all 
    fittings of this work.
    FWHM of spectral components (averaged  over magnetic field) corresponding to the
    data of Fig.~\ref{fig:Fig_B_dep} and 
    relaxation times $T_2^*$ calculated using Eq.~(\ref{eq:T2}).
    The values of  $\gamma_0$, $\gamma_a$ and $E_a$ are obtained from
    the fit of Eq.~\eqref{eq:act_low} to the temperature
    dependence shown in Fig.~\ref{fig:T_dep_spectra}(b).
    The value of $\sigma$ in the last line is calculated using Eq.~(\ref{eq:sigma}).
  }
  \label{tab:f0_FWHM}
  \begin{tabular}{|c|c|c|c|}
    \hline
    &electron&hole 1&hole 2\\
     \hline
    $g$-factor&$g_e=2.64$&$g_h=-0.39$&$g_h=-0.39$\\
    \hline
    $f_0$~(MHz)&$25\pm10$&$12\pm 5$&$35\pm 10$\\
    \hline
    $\sigma_0=f_0/\sqrt{2}$~(MHz)&$18\pm7$&$8.5\pm 3.5$&$25\pm 7$\\
    \hline
    FWHM~(MHz)&$45\pm 10$&$30\pm 10$&$90\pm 20$\\
    \hline
    $T_2^*$~(ns)&$7\pm 2$&$11\pm 4$&$3.5\pm 1$\\
    \hline
    $\gamma_0$~(MHz) &$7.2\pm4.9$&$4.6\pm 3$&$16.8\pm 3.4$\\
    \hline
    $\gamma_a$~(MHz) &$5000\pm2500$&$260\pm190$&$5000\pm 2500$\\
    \hline
    $E_a$~(meV) &$3.4\pm0.5$&$2.3\pm 0.9$&$4.9\pm 0.7$\\
    \hline
    $a_{loc}$~(nm) &$7.5\pm0.5$&$9.1\pm 1.8$&$6.2\pm 0.5$\\
    \hline
    $\sigma$~(MHz) from $a_{loc}$&$6\pm3$&$15\pm 4.5$&$26\pm 3$\\
    \hline
  \end{tabular}
\end{table}

\section{Probe energy   dependence}
\label{sec:probe_energy}
\begin{figure}
	\includegraphics[width=3.3in]{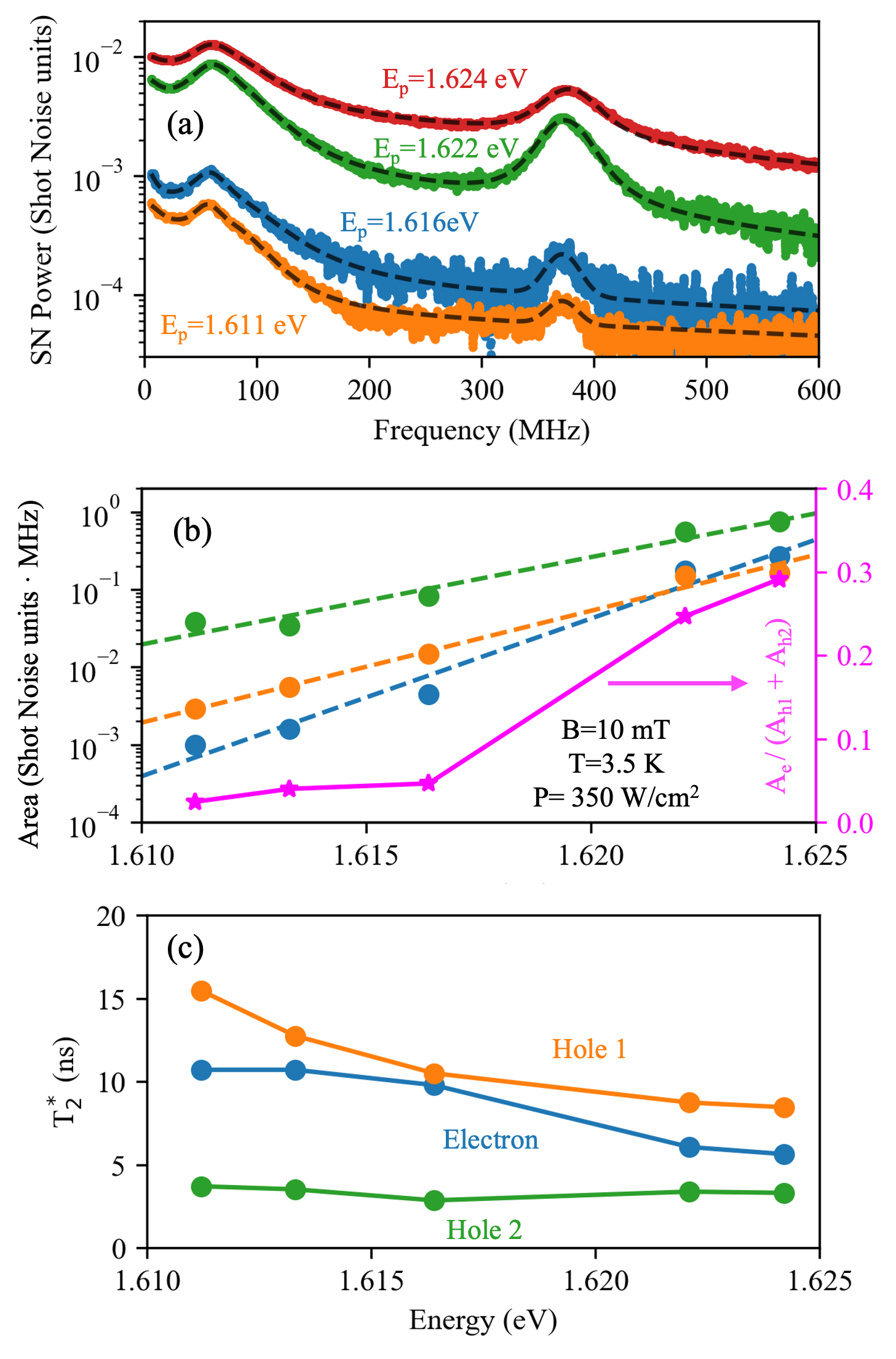}
	\caption{
		Probe energy dependence measured in forward-scattering geometry at 
		$T=3.5$~K, $P=350$~W/cm$^2$, $B=10$~mT.
		(a) SN spectra (symbols) and the results of the fitting (lines).
		(b) Areas of the electron 
		and hole contributions extracted from the fitting 
		procedure (symbols, left scale). 
		The corresponding linear fits (in logarithmic scale) are shown by dashed lines. 
		Electron contribution increases with energy faster 
		than the hole ones. The ratio between electron and 
		hole spectral areas is shown on the right scale. (c) Spin relaxation times $T_2^*$
		calculated according to Eqs.~\eqref{eq:FWHM} and ~\eqref{eq:T2}.
	}
	\label{fig:Fig_WL_analysis}
\end{figure}
The SN spectra measured in forward-scattering geometry at $T=3.5$~K, $P=350$~W/cm$^2$,
 $B=10$~mT and four different probe energies are shown in  Fig.~\ref{fig:Fig_WL_analysis}(a).
 These   energies are indicated on top of the PL spectrum, see Fig.~\ref{fig:setup}(b).
The spectrum at  the highest energy, $E_p=1.624$~eV, corresponds to vanishing transmission 
of the probe beam, the one at the lowest energy, $E_p=1.611$~eV, corresponds to vanishing SN. 
We base our  following analysis on the fitting function described in Sec.~\ref{sec:Identification}, 
with three Voigt profiles (the Gaussian parameter $\sigma$ is kept fixed, $\sigma=\sigma_0$, while $\gamma$ is allowed to vary) and two zero-frequency Lorentzian components, 
see dashed lines in Fig.~\ref{fig:Fig_WL_analysis}(a).

Figure~\ref{fig:Fig_WL_analysis}(b) shows how the spectral areas 
(symbols) associated with 
electrons and holes  depend on the probe energy.
One can see that both electron and hole spectral areas 
increase when energy increases. 
In the same time, the magenta symbols and curve show that the electron SN power increases with increase of the probe energy faster than that of both holes. 
%
We suggest that this is related to the relative positions of the 
corresponding trion resonances: the resonance of the negatively charged 
exciton lies below that of the positively charged exciton. 
As the probe energy increases, it therefore approaches the 
former resonance first, so the electron SN area grows 
faster than the hole one, even though the probe energy 
remains below both resonances at all probe energies.

Figure~\ref{fig:Fig_WL_analysis}(c) presents $T_2^*$ values 
extracted from the fits using Eq.~\eqref{eq:T2}. 
One can see that   spin relaxation times decrease with increase of energy. 
In principle, this can be related to acceleration of both nuclei-related ($\sigma$) and unrelated   ($\gamma$) contributions  with increase of the energy.
However, the experimental data do not allow us to tease apart the two mechanisms. 
Thus, as stated above,  for the sake of simplicity,   in the analysis shown in 
Fig.~\ref{fig:Fig_WL_analysis} we assumed  $\sigma=\sigma_0$.
We have checked that, leaving in the fitting procedure all the parameters free to vary
does not affect significantly the resulting values of $T_2^*$. 
Note also, that similar energy-dependence of $T_2^*$ has been reported 
for amplitudes and relaxation times of carriers 
contributing to time-resolved Faraday rotation in MAPbI$_3$ thin polycristalline films, 
although the measured times for electron were somewhat 
shorter~\cite{garcia-arellanoEnergyTuningElectronic2021}.  

\section{Probe power  dependence}
\label{sec:power}
\begin{figure}
	\includegraphics[width=3.3in]{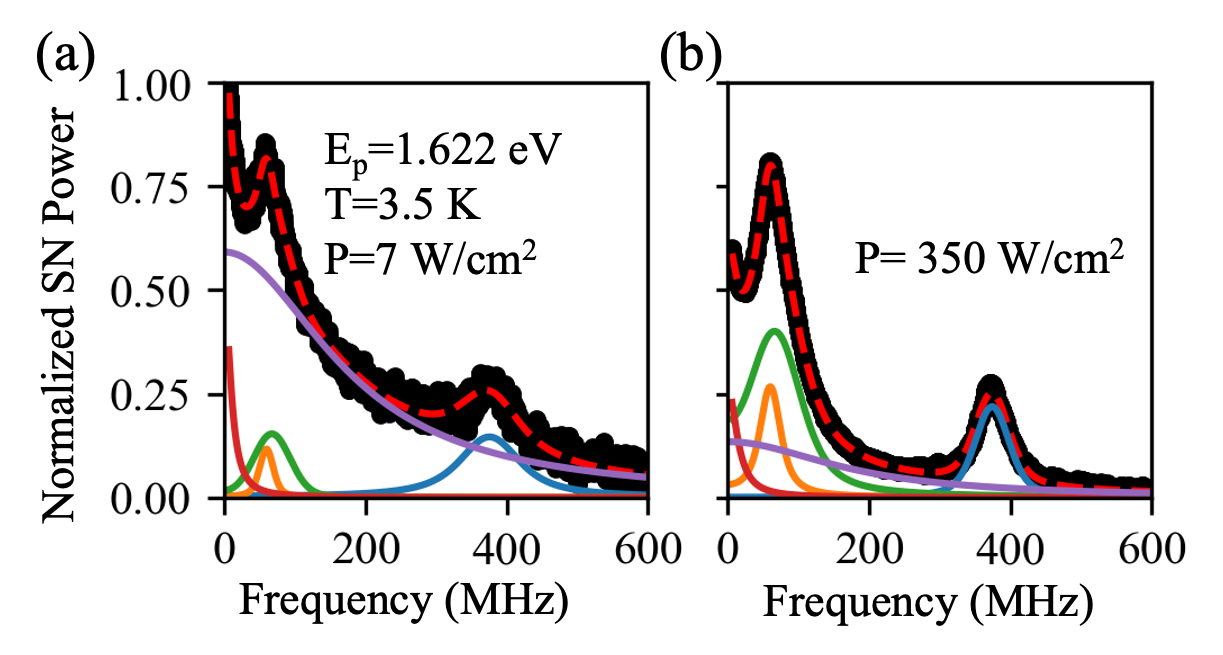}
	\caption{ Probe power dependence measured in forward-scattering 
	geometry at $T=3.5$~K, $E_p=1.622$~eV, $B=10$~mT:   
	SN spectra (black symbols)	 measured in forward 
	geometry at (a)  lowest power ($P= 7$~W/cm$^2$, 
	data multiplied by a factor of 20) and (b) highest  
	power ($P=350$~W/cm$^2$). 
	Red dashed line shows the fit to the multi-component 
	function with five components (solid lines), as described in Sec.~\ref{sec:Identification} .
	}
	\label{fig:P_dep_spectra}
\end{figure}

\begin{figure}
	\includegraphics[width=3.3in]{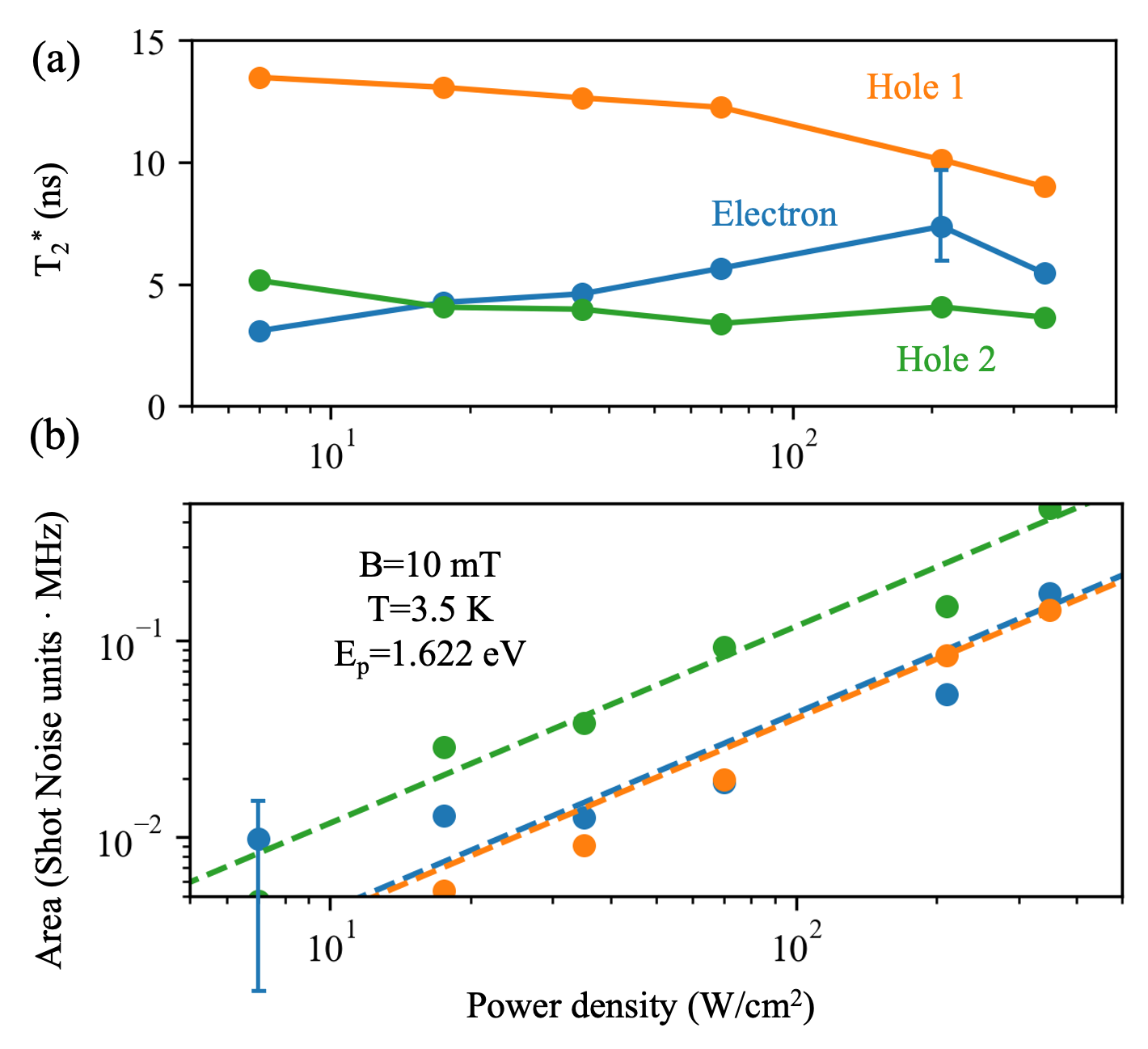}
	\caption{ Probe power dependence measured in 
	forward-scattering  geometry at $T=3.5$~K, $E_p=1.622$~eV, $B=10$~mT. 
	(a) Spin relaxation times 
	calculated after Eqs.~\eqref{eq:T2} and \eqref{eq:FWHM}. 
	(b)  Areas of the electron and hole contributions 
	extracted from the fits (symbols). Dashed lines are 
	 fits to linear dependence. 
	}
	\label{fig:P_dep_analysis}
\end{figure}

%
%
SN spectra measured at $E_p=1.622$~eV, $T=3.5$~K, $B=10$~mT
 and at two different power densities
in forward-scattering geometry are shown in Fig.~\ref{fig:P_dep_spectra},
 together with the fitting curves. 
Using the same  procedure of Voigt function fitting and determination 
of the FWHM as  in the previous section, we obtain the values of $T_2^*$ at  different powers 
shown in Fig.~\ref{fig:P_dep_analysis}(a). 
One can see that for holes spin relaxation time decreases with power.
This behavior is often observed in SN spectroscopy and time-resolved  Kerr rotation  
and interpreted as a result of either eventual heating of the system 
or exchange with photocreated carriers  \cite{kirsteinSpinDynamicsElectrons2022,Gribakin2025}.

For electron the situation is different and rather unusual: 
increasing the excitation power density leads to increase $T_2^*$
from $3$ to $7$~ns. 
This can be related to the increased density of photocarriers, 
which may influence indirectly the SN signal, either 
acting  to produce a kind of the motional narrowing effect 
on the exchange interaction between probed and surrounding spins,
or giving rise  to the  reduction of the localization volume of the 
electrons contributing in the SN signal. 
This is consistent with the linear dependence of the signal area on the probe power  shown in Fig.~\ref{fig:P_dep_analysis}(b), suggesting  that up to the highest powers used in this work, 
the SN probes  intrinsic, resident carriers. 
Nevertheless, the exact physics of the SN and spin relaxation of the electrons still needs to be  better understood in future studies.


\section{Temperature dependence}
\label{sec:temp}
\begin{figure}
	\includegraphics[width=3.3in]{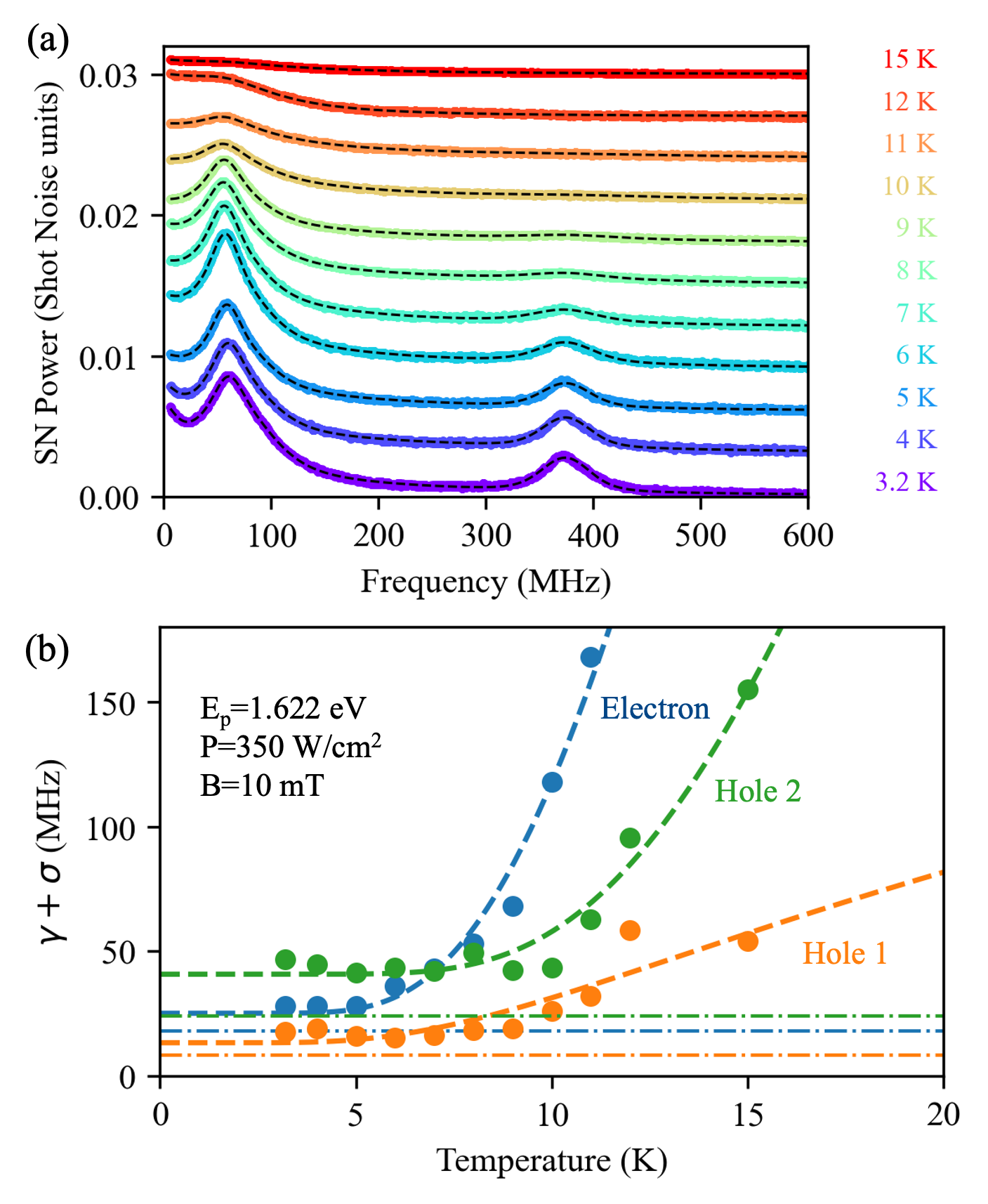}
	\caption{ 
		(a) Waterfall representation of SN spectra measured in 
		forward-scattering geometry at  temperatures from $3.5$~K to $15$~K, 
		$P=350$~W/cm$^2$, $E_p=1.622$~eV, $B=10$~mT.
		Black dashed lines show the results of the fitting procedure. 
		(b) Lorentzian  and Gaussian parameters 
		corresponding to  Voigt profiles extracted from  
		SN spectra shown in (a).
		Dashed-dotted lines show the Gaussian  parameter $\sigma_0$,
		which is assumed temperature-independent.
		Symbols represent  the sum of Lorentzian and 
		Gaussian  {contributions}  ($\gamma+\sigma_0$).
		Dashed lines are fits  to Eq.~\eqref{eq:act_low} with parameters 
		given in Table~\ref{tab:f0_FWHM}.
	}
	\label{fig:T_dep_spectra}
\end{figure}
Measuring and quantifying temperature dependence 
of the SN spectra is crucial to conclude on degree 
of carrier localization and on the role of various 
relaxation mechanisms \cite{sterinTemperaturedependentElectronSpin2022}. 
The temperature dependence of SN spectra is presented in 
Fig.~\ref{fig:T_dep_spectra}.
It is measured at $B=10$~mT, $E_p=1.622$~eV and $P=350$~W/cm$^2$
(the highest power density used in this work, {\it cf} Fig.~\ref{fig:P_dep_analysis}).
The corresponding lowest temperature spectrum is also shown in Fig.~\ref{fig:P_dep_spectra}(b).

One can see that with increasing temperature the SN signal fades out more rapidly for  electrons, 
than for holes, which is  consistent with previous observations \cite{kirsteinCoherentCarrierSpin2024}.
Previous studies established different possible scenarios for the temperature dependence of the spin noise spectra, that are based on the interplay between spin dynamics in the localized and excited (free) states. Typically, for localized carriers the Gaussian shape, 
characterized by parameter $\sigma$, dominates. Then activation of spin hopping into conduction or valence band (delocalization) leads to the increase of the Lorentzian 
width~\cite{Glazov_Ivchenko_2012,Glazov_hopping,Smirnov-review,zibinskiySpinNoiseLocalized2026}. 
In principle, spin hopping can also reduce the Gaussian contribution~\cite{sterinTemperaturedependentElectronSpin2022,Gribakin2025}. In our experimental data, there might be a weakly nonmonotonous dependence of the peaks widths on the temperature caused by the motional narrowing effect. 
However, its unequivocal evidence is difficult, so we leave its detailed investigation for the future studies. 

In what follows, we adopt a model developed recently for the spin noise of donor-bound electrons in CdTe quantum wells~\cite{zibinskiySpinNoiseLocalized2026}.
Specifically, we assume that the localized charge carriers can hop to the bands and back, or within the broad and shallow impurity bands, with a rate that grows with temperature increase. 
After each hopping event, the localization energy changes slightly.
Since both  probe laser  bandwidth and the homogeneous widths of the localized charged excitons are much smaller than the inhomogeneous broadening, this results in the removal of the given 
spin from the probed ensemble.  
Thus, the irreversible spin relaxation rate effectively increases, $\gamma(T) \ge \gamma(T=0) $. 
To account for this scenario in the fits, we fix $\sigma=\sigma_0$  (see Table~\ref{tab:f0_FWHM}) and allow $\gamma$ to vary.

The resulting parameters from the fits are shown in Fig.~\ref{fig:T_dep_spectra}(b). The values of $\sigma_0+\gamma$ are shown by symbols, while the values of $\sigma_0$ are shown  separately by dashed-dotted lines. 
One can see that at the lowest temperatures 
the values of $\sigma_0+\gamma$  saturate, while they increase with
increase of the temperature.
These dependencies 
can be  fitted assuming 
\begin{equation}
	 \gamma=	 \gamma_0 +\gamma_a \exp(-E_a/k_BT),
	\label{eq:act_low}
\end{equation} 
where $\gamma_0$ is the zero-temperature relaxation rate of the spin component parallel to the sum of the nuclear and external magnetic fields, $E_a$ is the activation energy and  $\gamma_a$ is the hopping prefactor~\cite{zibinskiySpinNoiseLocalized2026}. 
The activation 
energies $E_a$ can be interpreted as the binding energies 
in the model of shallow Coulomb trap states, and thus related to the carrier localization lengths 
$a_{loc}$ as ~\cite{PhysRevB.106.035202,Gribakin2025,Garcia-Arellano2019,zibinskiySpinNoiseLocalized2026}:
\begin{equation}
	a_{loc}=\hbar/\sqrt{2m_{e(h)}E_a}
\end{equation}
The resulting values of $\gamma_a$, $\gamma_0$, $E_a$ and $a_{loc}$
are summarized in Table~\ref{tab:f0_FWHM}, assuming $m_h=m_e=0.2m_0$, where
$m_0$ is the free electron mass \cite{parkVisualizingLowEnergyElectronic2024,umariRelativisticGWCalculations2014}. 

Let us now check whether these localization lengths are consistent 
with the values of $\sigma_0$, which, in 
this model, are directly related with the 
zero-temperature HF-induced  spin dephasing rates~\cite{book_Glazov}.
Since iodine contribution to HF interaction 
 for  holes is weak \cite{lagueHyperfineInducedAsymmetricOverhauser2026,KoturHyperfinePRB2026}, it can be neglected. 
 Assuming that only lead nuclei contribute to HF interaction, one expects ~\cite{zibinskiySpinNoiseLocalized2026}:
  \begin{equation}
    \label{eq:sigma}
    2\pi \sigma=\frac{A_{Pb}^s}{\hbar}\sqrt{\frac{I(I+1)xv_c}{24\pi a_{loc}^3}}.
  \end{equation}
Here, $I=1/2$ is the lead $^{207}$Pb  spin, $v_c=0.25$~nm$^3$ is the unit cell volume 
(this estimation is done for high symmetry cubic phase), $A_{Pb}^s=72~\mu$eV 
is the HF coupling constant~\cite{kirsteinSpinDynamicsElectrons2022,KoturHyperfinePRB2026,
lagueHyperfineInducedAsymmetricOverhauser2026}, 
and $x=0.22$ is the natural abundance of $^{207}$Pb.
Substituting these numbers, we obtain 
$ {\sigma^{h1}} =15\pm4.5$~MHz 
and $\sigma^{h2} =26\pm 3$~MHz.
These values are to be compared with the ones deduced from
 $f_0$, $ {\sigma_0^{h1}} =8.5\pm3.5$~MHz 
 and $\sigma_0^{h2} =25\pm 7$~MHz, shown by {dashed}-dotted lines in 
 Fig.~\ref{fig:T_dep_spectra}(b), used as input values in the fits
 shown in Fig.~\ref{fig:T_dep_spectra}(a), 
 and given explicitly in  Table~\ref{tab:f0_FWHM}, third line.
We consider the agreement as reasonable and conclude that 
 the experimental data consistently 
point on the HF-dominated hole spin dephasing.

For electrons the situation is completely different. Activation energy for the electron spin  $E_a^e=3.4\pm0.5$~meV, yielding  localization  length $a_{loc}^e=7.5\pm0.5$~nm.
Because HF interaction is known to be much weaker
for electrons than for holes,  one should expect   smaller values of $\sigma$ for electrons, {than for holes}.
On the basis  of the HF interaction theory in perovskites developed
in Refs~\cite{KoturHyperfinePRB2026,lagueHyperfineInducedAsymmetricOverhauser2026} and electron  localization length $a_{loc}^e=7.5\pm0.5$~nm  we obtain $\sigma^e=6\pm3$~MHz,  3 times less the value deduced from $f_0$ using Eq.~\eqref{sigma_0}, 
 $\sigma_0=18\pm 7$~MHz.
This discrepancy, as well as unconventional power dependence of  electron $T_2^*$ (see Sec.~\ref{sec:power}), suggest that mechanisms of the electron spin dephasing need to be reconsidered in future studies.

\section{Forward- versus back-scattering: \\
spatial distribution of spin carriers}
\label{sec:forwardback}
\begin{figure}
	\includegraphics[width=3.3in]{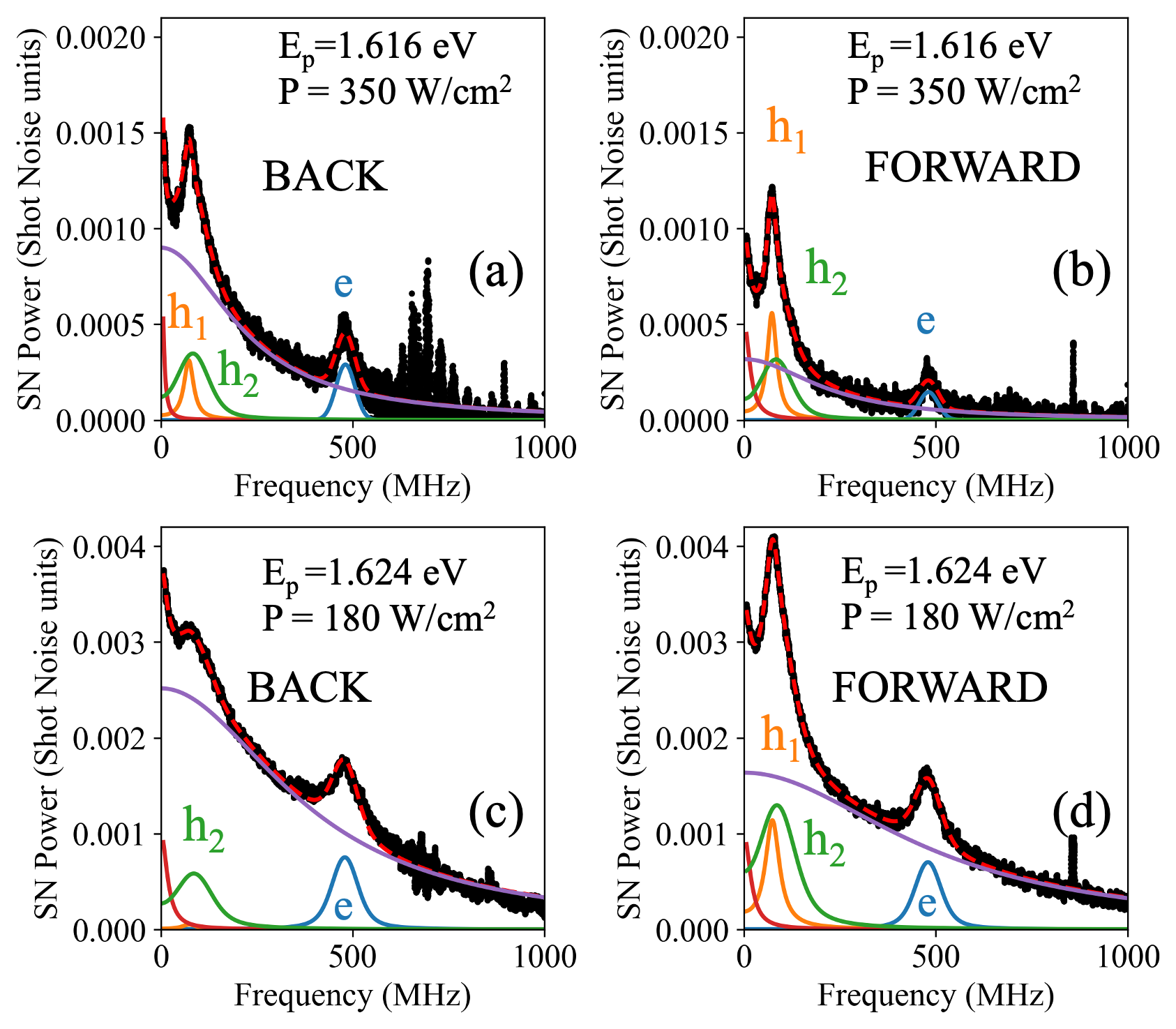}
	\caption{  Comparison between SN spectra (black symbols) 
	measured at $T=3.5$~K, $B=10$~mT  in forward-scattering (b, d) 
	and back-scattering (a, c) geometries.
	Measurements at two probe energies are shown: 
	(a-b) $E_p=1.616$~eV
	(weak absorption,  probe power $P=350$~W/cm$^2$) and (c-d) $E_p=1.624$~eV (strong absorption,  probe power 
	$P=180$~W/cm$^2$). 
	Red dashed lines show fits to Voigt profiles model, 
	assuming that back- and forward-scattering spectra 
	differ only by amplitude of the individual components, 
	while their central frequencies and widths do not change. 
	Solid lines represent the resulting individual 
	contributions of each component.
	}
\label{fig:BackForward}
\end{figure}
Combined  SN measurements in the forward- and back-scattering geometries (Fig.~\ref{fig:Fig1})  {can} provide information not only on the spin dynamics, 
but also on spin diffusion
\cite{Cronenberger2021,Cronenberger2019}.
Indeed, it was shown that spin diffusion among 
localization sites leads 
to broadening of the SN spectrum in back-scattering. 
Thus, one can expect to distinguish 
between localized and diffusing 
carriers using this approach.

Figure~\ref{fig:BackForward} shows comparison between SN spectra in different geometries at the two probe energies: $E_p=1.616$~eV and $1.624$~eV. 
The  spectra in forward-scattering geometry are fitted with five components, as described 
in Sec.~\ref{sec:Identification}. 
Then, to fit the spectra measured in back-scattering, only amplitudes of the components are allowed to vary, while the widths (both Gaussian and Lorentzian components) of all contributions are fixed.
Note, that the spectral widths that we measure in forward scattering depend  on  the probe energy, as discussed in Sec.~\ref{sec:probe_energy}, so that we fix them to these different values. 

The  assumption of identical linewidth  in the two detection geometries corresponds to the model of localized spins, meaning that they do not move during the spin lifetime.
Note however, that from the increase of the 
spectral width  that can be deduced  from the fits with free $\gamma$ values, 
we can deduce variations of FWHM, $\Delta_{\mathrm{FWHM}}$,  which 
do not exceed few tens of MHz. 
Thus, we can
estimate 
the diffusion coefficient:
\begin{equation}  
 {D = \frac{h^2 c^2}{16 \pi n^2 E^2_p}\Delta_{\mathrm{FWHM}}}\approx 0.5\times10^{-3}\mathrm{cm^2/s},
\label{eq:Dcoeff}
\end{equation}  
where $n\approx 2.7$ is the refractive index and $c$ is the speed of light.
This yields very small spin diffusion length  $l=\sqrt{DT_2^*}$ on the order of $1-2$ localization lengths, $a_{loc}$.

Still the amplitudes of the different contributions differ a lot between forward- and back-scattering   geometries, see Fig.~\ref{fig:BackForward}. 
In particular, the contribution from weakly localized hole (h$_1$) is extremely weak in the back-scattering geometry at the higher probe energy. 
Because the modifications of the spectrum   in  back-scattering with respect to forward-scattering geometry is weaker at lower probe energy, it seems reasonable to relate them to the absorption of the probe beam.

This effect can be described microscopically (see also Fog.~\ref{fig:Fig1}) by considering the light propagation along the $z$-axis through the sample of the length $L$ with the absorption coefficient $\alpha$. Then the amplitude of the light field scattered in  forward 
direction $\delta\bm{\mathcal E}_F$ by the spins $\bm S_i$ 
located at the coordinates $z_i$ has the form
\begin{equation}
  \delta\bm{\mathcal E}_F\propto\sum_i\e^{-\alpha z_i/2}S_{i,z}\e^{-\alpha(L-z_i)/2}=\e^{-\alpha L/2}\sum_iS_{i,z},
\end{equation}
where we sum up over all the spins in the sample. 
This means that the contributions of individual spins  to the amplitude of the scattered light  are  independent  on their positions with respect to the sample surface.
 By contrast, in the back-scattering geometry, the amplitude of light field has the form
\begin{equation}
  \delta\bm{\mathcal E}_B\propto\sum_i\e^{-\alpha z_i/2}S_{i,z}\e^{-\alpha z_i/2}=\sum_iS_{i,z}\e^{-\alpha z_i}.
\end{equation}
This means that the spins located closer to the surface contribute stronger to the back-scattered field, as expected in the presence of absorption, see Fig.~\ref{fig:Fig1}.

 The  spectral areas extracted from the fits of the spectra measured in  forward- and back-scattering geometries can not be compared directly, so we compare in Tab.~\ref{tab:Aratios} the ratios of different peaks areas between two measurement geometries. For example, at the higher energy, the ratio of areas of strongly and weakly localized holes, $A_{h2} / A_{h1}$, is $9$ times larger in back-scattering than in the forward-scattering. This indicates that h$_2$ are localized closer to the sample surface than h$_1$, as sketched in Fig.~\ref{fig:Fig1}. The same reasoning applied to electrons indicate 
that they are located even closer to the sample surface.   Overall, one can clearly see that the double  ratio
in Tab.~\ref{tab:Aratios}  differs from unity stronger at higher energy, where absorption is stronger,
as expected.

\begin{table}[h]
  \centering
  \caption{ {Ratios of relative areas of three contributions to the SN spectra between backward- and forward-scattering geometries for the two probe energies $1.616$~eV (LE) and $1.624$~eV (HE).}}
  \label{tab:Aratios}
  \setlength{\tabcolsep}{10pt} 
  \renewcommand{\arraystretch}{1} 
  \begin{tabular}{|c|c|c|c|}
    \hline
    Ratio&LE & HE \\
    \hline
    ${\left(A_{h2} / A_{h1}\right)_B}\: /\: {\left(A_{h2} / A_{h1}\right)_F}$&  $2$ &$9$ \\
    \hline
    ${\left(A_{h2} / A_{e}\right)_B}/{\left(A_{h2} / A_{e}\right)_F}$& $0.6$ &$0.4$ \\
    \hline
    ${\left(A_{h1} / A_{e}\right)_B}/{\left(A_{h1} / A_{e}\right)_F}$& $0.3$ &$0.04$ \\
    \hline
  \end{tabular}
\end{table}

It appears to be  difficult to analyze  these results more quantitatively, 
because the absorption coefficient varies by a 
factor of about 100 in the considered  
energy range \cite{dewolfOrganometallicHalidePerovskites2014} 
and because the form of all distributions functions is not known.  
Nevertheless we can hypothesize that, on average,
electrons are located most close to the surface,  weakly-localized holes 
with the longest spin dephasing times
are located in the bulk of the sample, while strongly localized 
holes are situated between them, as sketched in Fig.~\ref{fig:Fig1}.

\section{ Discussion and Conclusions}
\label{sec:concl}
Spin noise measurements, performed in forward- and back-scattering 
geometries on a MAPbI$_3$ single crystal, allow us to identify 
three families of intrinsic spin carriers: electrons and two groups 
of holes, h$_1$ and h$_2$, with different localization lengths, and thus 
different spin relaxation times. 
Identification of these carriers as intrinsic, rather than photocreated,
is made on the basis of linear (rather than quadratic)  power dependence 
of the spectral area associated with each of the components.

\paragraph*{Holes.} 
The observation of two hole contributions 
is in line with the Hanle and polarization recovery experiments 
on a similar sample \cite{KoturHyperfinePRB2026}, which suggests that 
spin noise and PL-based techniques probe the same families 
of carriers. 
We emphasize that the decomposition of the hole spectrum into exactly 
two components should be regarded as an effective description. 
Our data show that a single Voigt profile cannot reproduce 
the hole spin noise  {spectrum}, but they do not prove that the hole 
population is strictly bimodal.
 In a disordered crystal, 
the localization radii, and hence the number of nuclei 
within the localization volume and the HF field 
fluctuations, are expected to be broadly distributed. 
The resulting spectrum would then be a continuous superposition 
of lines with different widths and central frequencies, 
which two Voigt functions approximate reasonably well. 
h$_1$ and h$_2$ should thus be understood 
as two representative groups of holes, 
weakly and strongly localized, rather than as two well-defined species, 
and the parameters extracted for them 
($T_2^*$, $f_0$,  {$\gamma_0$}, $E_a$, $a_{loc}$) as average values characterizing 
the respective parts of a broader distribution. 
The probe power dependence indicates that we mainly probe 
resident holes spin dynamics, since the signal increases with power linearly.
The measured spin relaxation times are close to intrinsic ones, because at lowest powers they remain almost constant.
Comparison of back- and forward-scattering SN spectra indicates that 
holes are essentially localized.
Temperature, power and magnetic field dependencies of the SN 
spectra suggest that at low temperature spin relaxation and dephasing of both weakly and strongly localized holes are 
dominated by  HF coupling to nuclear spin system, and that DP mechanism is not 
relevant for hole spin dynamics in our 
sample, in agreement with conclusions of Refs.~\onlinecite{KoturHyperfinePRB2026} 
and \onlinecite{lagueHyperfineInducedAsymmetricOverhauser2026}.

Nevertheless, direct comparison between the Hanle widths measured in 
Ref.~\onlinecite{KoturHyperfinePRB2026} and our data is not straightforward.
Indeed, we observe  strong power and probe energy dependence of $T_2^*$,
and overall shorter relaxation times than the ones which can be deduced from 
Hanle widths. Moreover, the latter do not show any significant dependence on the detection energy.
Presumably, the state of the system under continuous wave excitation above exciton resonance 
modify the local spin environment  as compared to below-exciton resonance excitation used in
this work in a manner, that still need to be understood. 

\paragraph*{Electrons.} 
 {Electron spin properties deduced from SN measurements differ significantly from the hole ones.}  Surprisingly,  
electron spin relaxation time, $T_2^*$,  increases with increase of the probe power. 
Moreover, the Gaussian parameter, 
$\sigma$, estimated from the temperature dependence of the SN spectra is about three times 
smaller than the value $\sigma_0$ deduced from the zero-field 
frequency shift via Eq.~\eqref{sigma_0}.
 {These results suggest} that,
 {in contrast to holes},
HF interaction may not be the {main} mechanism limiting the electron 
spin dephasing, even at low temperature. 
A plausible candidate is the exchange interaction 
with other electrons or with holes, which should be more 
efficient for electrons than for holes, 
since the HF coupling 
of the former is weaker~\cite{lagueHyperfineInducedAsymmetricOverhauser2026,KoturHyperfinePRB2026}. 
As a consequence, the intrinsic electron spin dephasing time 
cannot be reliably extracted from our data.

\paragraph*{Spatial distribution of carriers.} 
Comparison of  forward- and back-scattering spectra at different 
probe energies shows that the relative contributions of the 
three carrier families depend on the absorption of the probe, 
while their widths and central frequencies remain 
essentially unchanged. 
Since forward-scattering is insensitive to 
the position of the scatterers, whereas back-scattering 
weights them with $\e^{-\alpha z}$, this reveals an inhomogeneous 
spatial distribution of carriers (Fig.~\ref{fig:Fig1}): 
electrons are located mostly in the vicinity of the crystal surface, 
weakly localized holes h$_1$ with the longest spin dephasing
times are found in the bulk, and strongly localized holes h$_2$ 
are situated between them. 
The observation of weakly localized holes in the bulk 
is consistent with their presence in single crystals and 
absence in thin films 
\cite{garcia-arellanoEnergyTuningElectronic2021,
	lagueHyperfineInducedAsymmetricOverhauser2026}.
The small variation of the  linewidth 
between the two geometries gives an upper bound for
diffusion coefficients, $D  < 0.5 \times 10^{-3}$~cm$^2$/s, 
confirming that carriers are not mobile on the time scale of their spin 
lifetime. 

\section{  Conclusions}
\label{sec:concl-concl}
In summary, spin noise spectroscopy of a MAPbI$_3$ single crystal 
reveals intrinsic electrons and holes, with hole spin 
relaxation at low temperature dominated by the HF 
interaction with nuclear spins. The localization energies, estimated 
from the temperature dependence of the spectra, are 
consistent with the measured  HF-induced dephasing rates 
and zero-field frequency shifts. The temperature dependence of the 
 linewidth is shown to be 
compatible with thermally activated hopping of localized holes  to the valence band.
 
The electron spin dynamics cannot be explained by 
 HF  interaction alone and is affected by photocarriers, 
possibly through exchange interaction. 
Measurements at different probe energies suggest that the localized negative trion resonance lies lower than that of the positive trion. Finally and most remarkably, the combination of 
forward- and back-scattering geometries shows that the 
charge carriers are distributed inhomogeneously across the crystal. 
These combined forward- and back-scattering SN measurements 
constitute a promising technique that can be applied to other materials.

Our results bridge, at least in part, the gap between spin noise 
and PL-based experiments, and call for measurements of 
the temperature dependence in both geometries on the same sample, 
as well as for a theoretical treatment of the exchange 
interaction and of the distribution of carrier localization lengths.

\section{Acknowledgements}
The authors thank D.~S.~Smirnov for his enormous contribution to the interpretation and understanding of the experiments and for many insightful discussions.
Financial support  from French National Research Agency, 
Grant No. ANR-21-CE30-0049 (CONUS) is gratefully acknowledged. 
D.R.Y. acknowledges the financial support by the Deutsche Forschungsgemeinschaft (Project YA 65/28-1, No. 527080192). The work at
ETH Zürich (B.T. and M.V.K.) was financially supported by the Swiss National Science Foundation (Grant
Agreement No. 200020E 217589) through the DFG-SNSF bilateral program and by ETH Zürich through ETH+ Project
SynMatLab.

\renewcommand{\i}{\ifr}
\bibliography{refs_cleaned}

\end{document}